# Impact of Intense Geomagnetic Storm on NavIC Signals Over Indore

Deepthi Ayyagari[(✉)], Sumanjit Chakraborty, Abhirup Datta, and Saurabh Das

Discipline of Astronomy, Astrophysics and Space Engineering, Indian Institute of Technology Indore, Simrol, Indore 453552, India
nagavijayadeepthi@gmail.com, {phd1701121001,phd1601121006, abhirup.datta,saurabh.das}@iiti.ac.in

**Abstract.** Intense geomagnetic storms can have a strong impact on the signals (termed as ionospheric scintillations) emitted by any global navigation satellite system (GNSS). The paper reports the first ever scintillations at Indore region on the NavIC signals due to impact of the intense geomagnetic storm event reported on September 8, 2017 at 01:51 and 13:04 UT. The variation of the planetary indices as well as the DST index which dropped to value of −124 nT on September 8, 2017 indicates the occurrence of an intense geomagnetic storm on September 8, 2017. The observations presented are carried out at Indore, which is located at the equatorial anomaly crest. The S4 index measurements of co-located GNSS receiver showed values ≥0.5 on the disturbed day between 15 and 18 UT. The analysis presented clearly signifies the degradation of carrier–noise measurements of NavIC L5 signal during the same time, which in turn affected the positional accuracy of NavIC, an important consideration for performance.

## 1 Introduction

Ionosphere, as the name indicates is the ionized layer of the atmosphere roughly stratified between 50 and 1000 km of altitude from the surface of the Earth, formed by the interaction of sunlight with various gases that aid the formation of the atmosphere [1]. With the advent of technology, many global navigation satellite system (GNSS) aid in positioning precision, remote sensing and other applications for the users. Ionospheric delay (iono-delay) induced in the signals transmitted by the GNSS is the major source that affects the accuracy in these applications, and this effect of the ionosphere depends on the frequency of signal transmitted and the refractive index of the medium. The iono-delay measurements are inversely proportional to the frequency of signal transmitted by GNSS along the path from satellite to receiver [2]. In addition, high frequency (HF) signal propagation from GNSS is also affected by ionospheric irregularities. The two most important effects are increase in ionization of E and D regions and the plasma instabilities. The former effect increases the collision frequency, which accelerates radio wave absorption results in loss of signal strength and triggers radio blackouts. In the E and F region, ionosphere irregularities can alter the reflection height, induce scattering and change the direction of propagation of the signal which again results in loss of signal. The change in direction of propagation of signal plays a major role in navigational accuracy [3]. The latter effect is also a source of ionospheric irregularities but has its significance at high latitudes [4, 5].

## 2 Ionospheric Scintillation

Ionospheric scintillation is a phenomenon which is best described as a rapid or sudden change in phase and amplitude of the GNSS signal when it passes through ionosphere [6]. Studies reveal that effects of ionospheric scintillation are more severe in the equatorial and low-latitudes regions, as well as in the high latitudes regions [7]. Ionospheric scintillation events in high latitudes are usually related to periods of high solar activity and geomagnetic storms. In the equatorial and low-latitudes regions, such events occur mainly due to equatorial ionization anomaly (EIA) and ionospheric bubbles (equatorial plasma bubbles (EPB)), formed in this region post-sunset [8]. Intense EPB formation leads to severe scintillation in the radio wave communications from satellites to ground. This in turn degrades the accuracy in determining the position by global navigation satellite systems (GNSS). For the horse latitudes (45°) and beyond regions, the scintillation irregularities have been well characterized in the literature.

Some quantitative information of the scintillation is measured as $S_4$ and $\sigma_\varphi$ indexes [9], it is possible to characterize and understand the ionospheric irregularities that cause ionospheric scintillations with these indices. The *S*-value is a way to characterize the power variation of a signal as a function of time [10, 11]. The $S_4$ is most widely used index among all the *S* indices and aids in mapping the intensity of ionospheric scintillation. The $\sigma_\varphi$ index indicates the carrier phase measurement ($\varphi$) and its variation at the receiver during the past 60 seconds, quantizing the standard deviation of the GPS signal phase. The $S_4$ index is defined as the normalized variance of intensity of the signal [12, 13] and is given as

$$S_4^2 = \frac{\langle I^2 \rangle - \langle I \rangle^2}{\langle I \rangle^2} \tag{1}$$

where *I* is the intensity of the signal. The *S*4 index can be classified to three categories depending directly on the signal intensity that occurred on the day and place and are given as strong ($S4 \geq 0.6$), moderate ($0.3 \leq S4 \leq 0.6$) and weak ($0.1 \leq S4 \leq 0.3$) [14].

## 3 Present Study

The Discipline of Astronomy, Astrophysics and Space Engineering (DAASE) of Indian Institute of Technology, Indore (Lat: 22.52° N, Lon: 75.92° E) operates a dual frequency NavIC, from May 2017, provided by Space Applications Centre, ISRO, capable of receiving GPS L1, NavIC L5 and *S*1 signals. Apart from the NavIC receivers, the discipline is well equipped with another multi-frequency and multi-constellation GNSS (GPS (L1,

L2 and L5), GLONASS(G1, G2 and G3) and GALILEO (E1, E5, E5a, E5b,E6)) receiver which is operational from June 2016 whose ionospheric pierce points (IPP) is shown in Fig. 1. The present study utilizes carrier–noise ratio (*C*/No) data from NavIC receiver measured as (dB-Hz), and the scintillation indices estimated for all GPS operational frequencies $S4$ to analyse the signal strength of NavIC during the disturbed period of the ionosphere. As the receivers are located at Indore, geographically located at the crest of equatorial ionization anomaly, analysis based on the data of such locations give the best estimate of the potential of the signal strength of these receivers. An event reported by Space Weather Prediction Centre [15] for the arrival of Coronal Mass Ejection (CME) on September 6, 2017, which lasted till September 7, 2017 triggered a G4 level geomagnetic storms at 23.50 UT, on September 7, 2017 as well as September 8, 2017 at 01:51 UT and at 13:04 UT. The variation of the planetary indices [16] as well as the DST index [17] which dropped to value of $-124$ nT on September 8 2017 indicates the occurrence of an intense geomagnetic storm on September 8, 2017. However, on any geomagnetically quiet day measurements of *C*/No of NavIC L5 signal and the $S_4$ index would remain as represented in Fig. 2. The values of *C*/No (as shown in Fig. 2) of all the satellites of NavIC range from 43 to 52 dB-Hz on any geomagnetically quiet day. The corresponding $S_4$ index values given by GPS are clearly below the value of 0.2 which indicate a very weak scintillation, not strong enough to disturb the strength of signal.

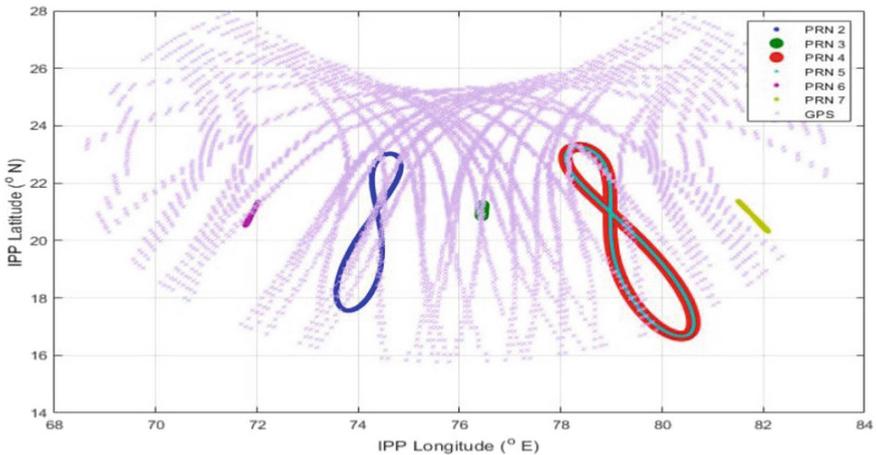

**Fig. 1** IPP (tracks) of the satellite vehicles of NavIC and GPS

## 4 Results and Conclusion

On September 8, 2017 due to the effect of intense geomagnetic storm, the *C*/No value of NavIC L5 signal of PRN 5 and 6 shrunk below 40 dB-Hz during the time of storm *TS* around 15–18 UT and the corresponding $S4$ value peaked beyond the mark of 0.6 (area marked in ellipse of Fig. 3) during the same time significantly indicating the storm impact over the signal strength of NavIC satellite system. Moreover, the accuracy in estimating the position coordinates (latitude, longitude and altitude) has degraded significantly and is shown in Fig. 4, and the data of position coordinates has been sampled to two parts during the *Ts*, i.e., 15–18.

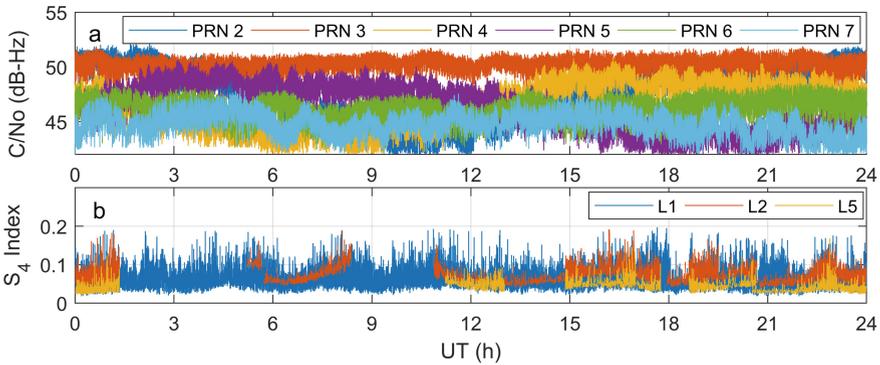

**Fig. 2** **a** *C*/No values of all the NavIC satellites and **b** *S*4 index from GPS satellites as a function of time for one day

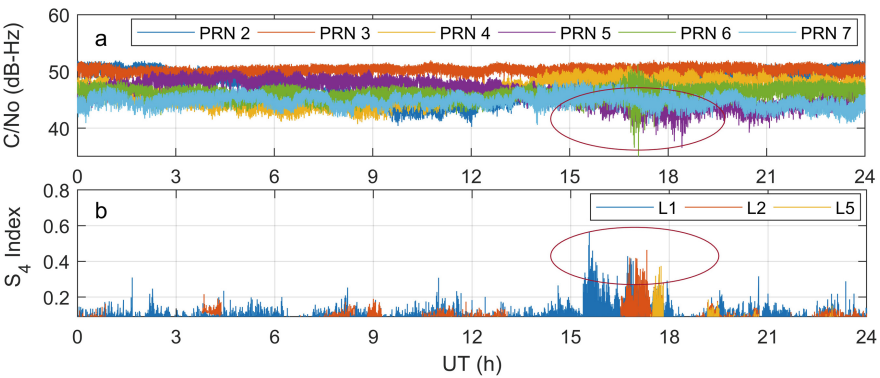

**Fig. 3** **a** *C*/No values of all the NavIC satellites and **b** *S*4 index from GPS satellites as a function of time for one day. The region marked with ellipses clearly signifies the impact of scintillation over NavIC L5 signal

UT and before the *Ts* during 3–6 UT, then the distribution of the sample points is shown in Fig. 5a–c. The error in the position accuracy of latitude, longitude and altitude is observed to be 1.0273 m, 1.5789 m and 1.9905 m, respectively. The spread of points is clear proof that the scintillation-induced storm clearly affects the accuracy of position estimates, which in turn signifies the impact of geomagnetic storms on NavIC signals. Furthermore, such studies should be carried out throughout the Indian region in order to investigate the effects of scintillation triggered due to impact of intense storm. Such studies would help formulating modelling techniques to detect such scintillation events on signals and on the methods to mitigate these effects.

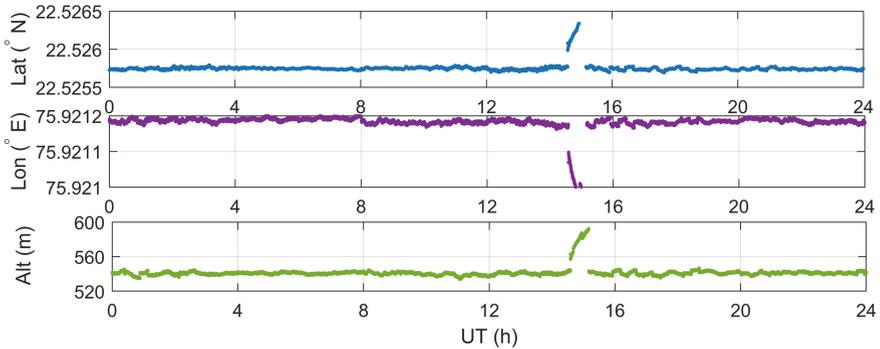

**Fig. 4** Plots represents the degradation in the estimates of position coordinates, i.e., latitude (Lat), longitude (Lon) and altitude (Alt) on September 8, 2017

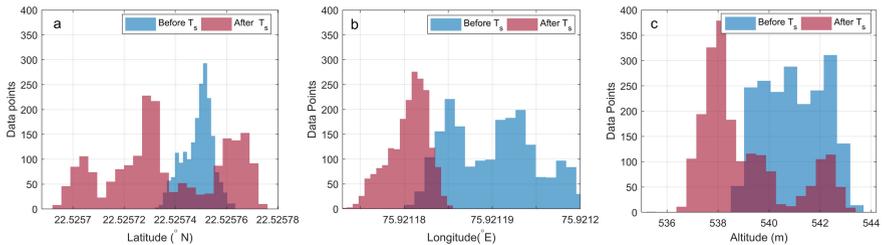

**Fig. 5** Each of the plot (**a** latitude, **b** longitude, **c** altitude) represents the distribution of the samples of position estimates before and after the affect of scintillation induced in the signals of NavIC system due to the impact of storm

**Acknowledgements.** Deepthi acknowledges Department of Science and Technology (DST) for providing her the DST-Inspire fellowship grant to pursue her research. The authors acknowledge Space Applications Centre (SAC), ISRO for providing the NavIC data (ACCORD receiver) under the project number: NGP-17 to Discipline of Astronomy, Astrophysics and Space Engineering, IIT Indore.